\documentclass[letter,traditabstract]{aa}
\usepackage{graphicx}
\usepackage{txfonts}
\usepackage{subfigure}

\begin{document}

\title{Water emission from the high-mass star-forming region IRAS\,17233-3606\thanks{{\it Herschel} is an ESA space observatory with science instruments provided by European-led Principal Investigator consortia and with important participation from NASA.}}\subtitle{High water abundances at high velocities}
   \author{S.\,Leurini
          \inst{1}
          \and A.\,Gusdorf\inst{2}
          \and F.\,Wyrowski\inst{1}
          \and C.\,Codella\inst{3}
          \and T.\,Csengeri\inst{1}
          \and F.\,van der Tak\inst{4,5} 
          \and H.\,Beuther\inst{6}
          \and D.\,R.\,Flower\inst{7}
          \and C.\,Comito\inst{8}
          \and P.\,Schilke\inst{8}}
   \offprints{S.\,Leurini}

\institute{Max-Planck-Institut f\"ur Radioastronomie, Auf dem H\"ugel 69, 53121 Bonn, Germany, 
\email{sleurini@mpifr.de}
\and LERMA, UMR 8112 du CNRS, Observatoire de Paris, \'Ecole Normale Sup\'erieure, 24 rue Lhomond, F75231 Paris Cedex 05, France
       \and INAF - Osservatorio Astrofisico di Arcetri, Largo E. Fermi 5, 50125 Firenze, Italy
\and SRON Netherlands Institute for Space Research, PO Box 800, 9700 AV, Groningen, The Netherlands
\and Kapteyn Astronomical Institute, University of Groningen, PO Box 800, 9700 AV, Groningen, The Netherlands
\and Max-Planck-Institute for Astronomy, K\"onigstuhl 17, 69117, Heidelberg, Germany
\and Physics Department, The University, Durham DH1 3LE, UK
\and Physikalisches Institut, Universit\"at zu K\"oln, Z\"ulpicher Str. 77, 50937 K\"oln, Germany}
\date{\today}

\abstract{We investigate the physical and chemical processes at work during the formation of a massive protostar based on the observation of water in an outflow from a very young object previously detected in H$_2$ and SiO in the IRAS 17233--3606 region. We estimated the abundance of water to understand its chemistry, and to constrain the mass of the emitting outflow. We present new observations of shocked water obtained with the HIFI receiver onboard \textit{Herschel}. We detected water at high velocities in a range similar to SiO.
We self-consistently fitted these observations along with previous SiO data through a state-of-the-art, one-dimensional, stationary C-shock model. We found that a single model can explain the SiO and H$_2$O emission in the red and blue wings of the spectra.
 Remarkably,  one common area, similar to that found for  H$_2$ emission,  fits both the SiO and H$_2$O emission regions. This shock model subsequently allowed us to assess the shocked water column density, $N_{\rm H_2O}=1.2\,10^{18}$~cm$^{-2}$, mass, $M_{\rm H_2O}=12.5~M_\oplus$, and its maximum fractional abundance with respect to the total density, $x_{\rm H_2O}=1.4\,10^{-4}$.  The corresponding water abundance in fractional column density units ranges between $2.5\,10^{-5}$ and $1.2\,10^{-5}$, in agreement with recent results obtained in outflows from low- and high-mass young stellar objects.}


   \maketitle
%

\section{Introduction}

The formation mechanism of high-mass stars ($M>8$\,$M_\odot$) has
been an open question despite active research for several decades now, the main reason being
that the strong radiation pressure exerted by the young massive star
overcomes its gravitational attraction
\citep{1974A&A....37..149K}. Controversy 
remains about how high-mass young stellar objects (YSOs)  acquire their mass \citep[e.g., ][]{2009sfa..book..288K},
either locally in a prestellar phase
 or  during the star formation process itself, being funnelled to the centre of a stellar cluster by the cluster's
gravitational potential. Bipolar outflows are a natural by-product of  star
formation and understanding them can give us important insights into the way massive stars form. 
In particular, studies of their properties in terms of morphology and
energetics as function of the luminosity, mass, and evolutionary phase
of the powering object may help us to understand whether the
mechanism of formation of low- and high-mass YSOs is the same or
not \citep[see, e.g.,][]{2002A&A...383..892B}.

Water is a valuable tool for  outflows as it is
predicted to be copiously produced under the type of shock conditions
expected in outflows \citep{Flower10}. Observations  of molecular outflows powered by
YSOs of different masses
 reveal  abundances of H$_2$O associated with outflowing gas of the order of some 10$^{-5}$ 
  \citep[e.g., ][]{2010A&A...521L..28E,2012A&A...542A...8K,2013A&A...549A..16N}. Recently, the Water In
Star-forming regions with Herschel \citep{2011PASP..123..138V} key program targeted several
 outflows from Class 0 and I low-mass YSOs in water lines. 
  H$_2$O emission in  young Class 0 sources is dominated by outflow components; in Class I YSOs H$_2$O emission is weaker because of less energetic outflows \citep{2012A&A...542A...8K}. 
 Comparisons of low-excitation water data with SiO, CO, and H$_2$ reveal contrasting results because
these molecules seem to trace different environments in some sources 
 \citep{2013A&A...549A..16N,2013A&A...551A.116T} while they have similar profiles and morphologies in others \citep{2012ApJ...757L..25L,2012A&A...538A..45S}.
 Observations of massive YSOs \citep[e.g.,][]{2013A&A...554A..83V}
confirm broad profiles due to outflowing gas in low-energy H$_2$O lines.  However, the coarse spatial resolution of
{\it Herschel} and the limited high angular resolution complementary
data   resulted in a lack
of specific studies dedicated to outflows from massive
YSOs.

\begin{figure}
\centering
\includegraphics[angle=-90,width=7cm]{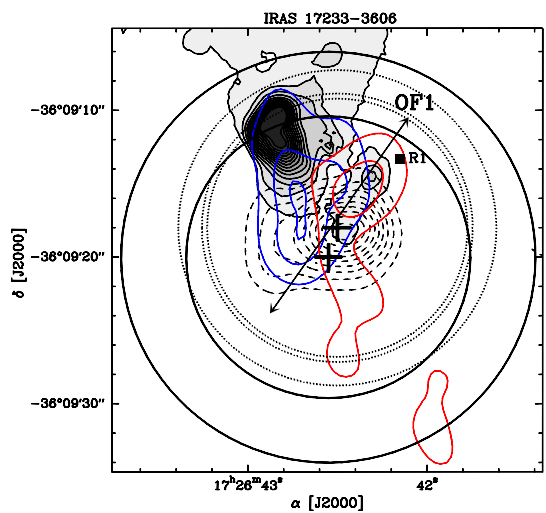}
\caption{Grey scale and solid black
contours represent the H$_2$ emission at 2.12$\mu$m; dashed contours are the 1.4 mm continuum emission. Red and blue contours are the SMA integrated emission of the SiO(5--4) line ( $\varv_{\rm{bl}}=[-30,-20]$ km\,s$^{-1}$ and $\varv_{\rm{rd}}= [+10,+39]$ km\,s$^{-1}$). The  crosses mark the {\it Herschel} pointings;  the solid and dotted circles are the {\it Herschel} beams (Sect.\,\ref{obs}). The square marks the peak of the EHV CO(2--1) red-shifted emission (R1). The arrow marks the OF1 outflow.}\label{overview}
\end{figure}

The prominent far-IR source IRAS\,17233$-$3606 (hereafter IRAS\,17233) is one of the best
laboratories for studying massive star formation because of its close
distance \citep[1\,kpc,][]{2011A&A...530A..12L}, high luminosity,
 and  relatively simple geometry. In 
previous interferometric studies, we resolved three CO outflows with high
collimation factors and extremely high velocity  (EHV) emission \citep[][Paper\,I]{2009A&A...507.1443L}. Their kinematic ages ($10^2-10^3$ yr)  point to deeply embedded YSOs that still have not reached
the main sequence. 
One of the outflows, OF1 (Fig.\,\ref{overview}),
was the subject of a dedicated analysis in SiO lines
\citep[][Paper\,II]{2013A&A...554A..35L}. It is associated with EHV
CO(2--1), H$_2$, SO, and SiO emission. SiO(5--4) and (8--7)
APEX spectra suggest an increase of  excitation with velocity and point to hot and/or dense gas close to the primary jet. Through a combined 
shock-LVG analysis of SiO, we derived a mass of
$>0.3\,M_\odot$ for OF1, which implies a luminosity L$\ge10^3\,L_\odot$ for its driving source.

In this Letter, we present observations of water  towards IRAS\,17233 
with the HIFI instrument \citep{2010A&A...518L...6D} onboard {\it
  Herschel} \citep{2010A&A...518L...1P}.

\section{Observations}\label{obs}

Six water lines and  one H$_2^{18}$O transition were observed towards the  positions $\alpha_{\rm{J2000}}=17^{\rm h}26^{\rm m}42\fs50$, $\delta_{\rm{J2000}}=-36\degr09\arcmin18\farcs00$ (OBSIDs 1342242862, 1342242863, and 1342242875),  and 
$\alpha_{\rm{J2000}}=17^{\rm h}26^{\rm m}42\fs54$, $\delta_{\rm{J2000}}=-36\degr09\arcmin20\farcs00$ 
(OBSIDs 1342266457 and 1342266536) with a relative offset of ($0\farcs5,-2\farcs0$).   
Conversion to $T_{\rm{mb}}$ was made using the beam efficiencies
given in Table\,\ref{table1} and a forward efficiency of 0.96. Data were taken simultaneously in H and V polarisations using  
the acousto-optical Wide-Band Spectrometer.
OBSIDs 1342242862, 1342242863, and 1342242875 were acquired in spectral scan mode with a redundancy of 4  to allow for sideband separation \citep{2002A&A...395..357C}. The data were calibrated with the standard calibration pipeline within HIPE\,11.0 \citep{2010ASPC..434..139O}. Sideband separation was performed using the GILDAS\footnote{http://www.iram.fr/IRAMFR/GILDAS} CLASS package.
OBSIDs 1342266457 and 1342266536 were taken in single-pointing mode and level\,2 data were exported  into CLASS90 where they were analysed in detail. After inspection, 
data from the two polarisations were averaged together.

\section{Observational results}\label{obs_res}
\begin{figure}
\centering
\includegraphics[bb = 72 27 568 755,clip,width=8.5cm]{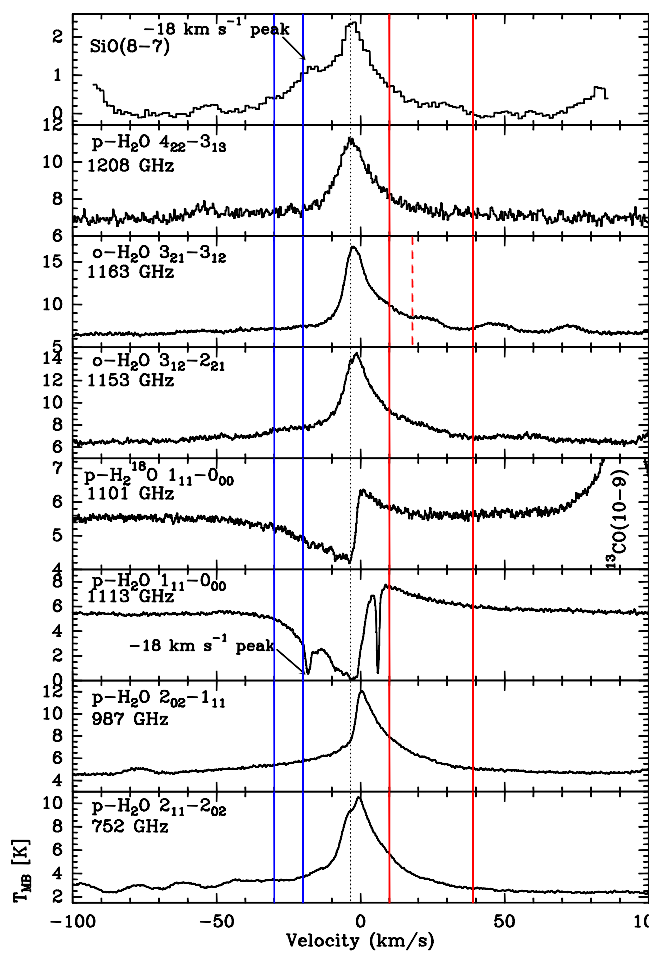}
\caption{Spectra of the water lines and of the SiO(8--7) transition towards
  IRAS\,17233--3606.  In all {\it Herschel} spectra, the continuum
  level is divided by a factor of two to correct for the fact that
  HIFI operates in double-sideband.  The red and blue lines mark the
  velocity range  used for the modelling of the
  water emission (solid red: [+10,+39]\,km\,s$^{-1}$; dashed red: [+18]\,km\,s$^{-1}$; blue: $[-30,-20]$\,km\,s$^{-1}$, Sect.\,\ref{model}).
The SiO(8--7) spectrum has
  a beam size of 18\arcsec, similar to the 19\arcsec\ beam size of
  the 1113\,GHz line, and it is observed at (4\farcs7,0\farcs0) from the \textit{Herschel} $\alpha_{\rm{J2000}}=17^{\rm h}26^{\rm m}42\fs54$, $\delta_{\rm{J2000}}=-36\degr09\arcmin20\farcs00$ pointing. The dotted line
  marks the ambient velocity.}\label{h2o} 
\end{figure}

Figure\,\ref{h2o} shows the H$_2$O spectra towards
IRAS\,17233. 
In all transitions, we detected water at high-velocities with respect to the ambient velocity \citep[$\varv_{\rm{LSR}}=-3.4$\,km
  s$^{-1}$,][]{1996A&AS..115...81B}: indeed, IRAS\,17233 presents one of the broadest profiles in the $1_{11}-0_{00}$ transition in high-mass YSOs \citep{2013A&A...554A..83V} known to date.
The ground-state line shows narrow absorptions at 
 $-18$ and $+6$\,km\,s$^{-1}$. They might be due to different clouds along the line of sight.  
However, 
the SiO(8--7) line, observed 
with a similar angular resolution (Paper\,II), has a well-defined emission peak at  $-18$\,km\,s$^{-1}$ (see
Fig.\,\ref{h2o}) although broader than the H$_2$O absorption. At  $-18$ and $+6$\,km\,s$^{-1}$ \citet{2008AJ....136.1455Z} detected H$_2$O maser spots coming from  the region shown in Fig.\,\ref{overview}. These absorptions
might be due to cold water associated with the outflows. 
The H$_2$O and H$_2^{18}$O ground-state lines
have 
deep blue-shifted absorptions against the continuum and the outflow at
velocities up to $-50$\,km\,s$^{-1}$,
 while the main isotopologue line
shows red-shifted emission up to $50$\,km\,s$^{-1}$ and its H$_2^{18}$O
equivalent up to $+17$\,km\,s$^{-1}$.   High-velocity red-shifted emission is detected up
to +50/60\,km\,s$^{-1}$ in all other lines, exept in the highest energy line ($p$-H$_2$O $4_{22}-3_{13}$) where emission is detected only up to $+18$\,km\,s$^{-1}$. 
The red-shifted wing of the 1163\,GHz line and the blue-shifted wing of the 752\,GHz transition are contaminated by
hot-core-like features. Emission up to $-70$\,km\,s$^{-1}$ is detected in the other transitions.

Comparison of the H$_2$O and H$_2^{18}$O  $1_{11}-0_{00}$ profiles in
the red-wings shows that the main isotopologue line is deeply
affected by absorption also at high velocities since red-shifted
emission is detected from 1.3\,km\,s$^{-1}$ in H$_2^{18}$O and only
from 9\,km\,s$^{-1}$ in H$_2$O (Fig.\,\ref{h2o}).  The line ratio between the two
$1_{11}-0_{00}$ isotopologue lines ranges between 0.95 and 0.3 in the
blue wing ($[-30,-20]$\,km\,s$^{-1}$), establishing very high opacities for the
main isotopologue transition even at high velocities and suggesting that 
it  may be contaminated
by a component in emission. Indeed, assuming negligible excitation with respect to the continuum, the opacity
of the H$_2^{18}$O line is between 0.02 and 0.3 in the velocity
interval $[-50,-4]$\,km\,s$^{-1}$ (see Eq.\,1 \,of
\citealt{2012A&A...542A..76H}). This corresponds to a column density
of H$_2^{18}$O of $8.4\,10^{11}$\,cm$^{-2}$ at the peak of the
absorption, down to $5.5\,10^{10}$\,cm$^{-2}$ in the high-velocity
wing ($-50$\,km\,s$^{-1}$). The total $p$-H$_2^{18}$O column density over the velocity range $[-50,-4]$\,km\,s$^{-1}$ is $1.2\,10^{13}$\,cm$^{-2}$. Assuming that the 1113\,GHz thermal continuum has the same distribution as at 1.4\,mm  
(deconvolved size at $FWHM$ of   
$5\farcs3\times2\farcs7$, Paper\,I and Fig.\,\ref{overview}), we corrected the continuum emission for beam dilution in the {\it Herschel} beam (Table\,\ref{table1}) and estimate
 a $p$-H$_2^{18}$O column density of $2.4\,10^{14}$\,cm$^{-2}$, which corresponds to a  total column density of H$_2$O of
$5.3\,10^{17}$\,cm$^{-2}$ for a standard isotopic ratio
$^{16}\rm{O}/^{18}\rm{O}=560$ \citep{1994ARA&A..32..191W} and a ortho-to-para ratio of 3. This is most likely a lower limit to the H$_2$O column density since the 1113\,GHz thermal continuum is probably more compact than that at 1.4\,mm.

 Given the complexity of the 1113\,GHz line at low-velocities, we focussed our  analysis on the outflow component detected at high-velocities. 
The similarity of the SiO and H$_2$O profiles
suggests a common origin of the high-velocity emission in the two
molecules. 
Therefore, we limited our  analysis to the velocity ranges $[+10,+39]$\,km\,s$^{-1}$ and $[-30,-20]$\,km\,s$^{-1}$ used in Paper\,II.  For the 1163\,GHz line, we used the velocity range $[+10,+18]$ km\,s$^{-1}$. We did not include the 752\,GHz blue wing in the analysis because of severe contamination from other features.

\section{Shock-model of  the water emission}\label{model}
\begin{figure}
\centering
\includegraphics[width=0.45\textwidth]{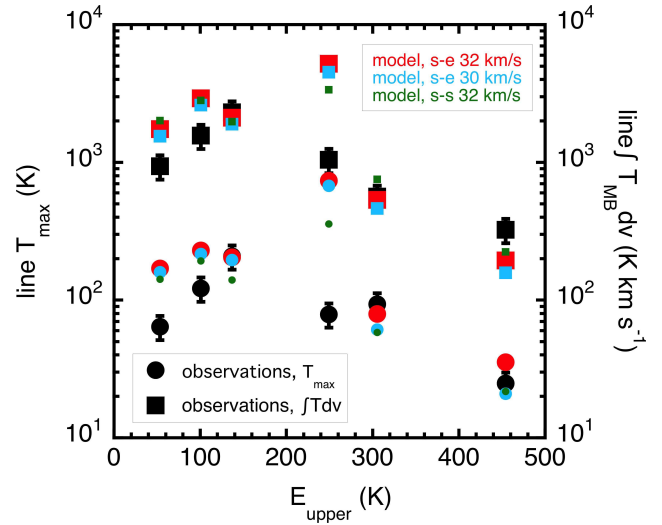}
\caption{Observed and modelled maximum brightness temperatures (circles), and integrated intensities (squares) for the red lobe of OF1. Data (in black) are corrected for an area of 6\,arcsec$^2$, and for 60\% of the emission due to OF1. Error bars are $\pm$20\% of the observed values. Three models are shown: the model of Paper\,II with level populations in statistical equilibrium (\lq s-e' in red) with $\varv_{\rm s}=32$\,km\,s$^{-1}$, one with a slower shock velocity ($\varv_{\rm s}=30$\,km\,s$^{-1}$, blue), and a model in stationary-state (\lq s-s' in green).}
\label{H2O-final-2in1}
\end{figure}

In Paper II, we demonstrated that the SiO emission in OF1 can be reproduced by a C-type shock model. We interpreted the SiO (8--7) and (5--4) emission at high velocities as due mostly ($\sim$60\%) to the OF1 outflow and modelled  their maximum brightness temperature and wing-integrated line ratio. Our best fit was found for a pre-shock density $n_{\rm H}=10^6$\,cm$^{-3}$, shock velocity $\varv_{\rm s}=32$\,km\,s$^{-1}$, magnetic field strength $B = 100$\,$\mu$G, and an age between 500 and 1000\,yr, in agreement with observations (Paper I). The emitting area of the SiO (5--4) transition is similar to that of H$_2$, 6\,arcsec$^2$, with an upper limit of 22\,arcsec$^2$. Our goal is to determine if the SiO-fitting shock can also reproduce the observed H$_2$O emission. Since the SiO modelling was performed towards a position $\sim9\arcsec$ off from the \textit{Herschel} pointing, our first step was to verify that the model of Paper II is also valid on this position. We then post-processed the shock model with an LVG module to calculate the radiative transfer of water lines \citep{Gusdorf11}. We thus compared modelled maximum brightness temperatures and integrated intensities to their observed values for two lines of o-H$_2$O and four lines of p-H$_2$O, under the exact same assumptions as adopted  for SiO: emitting area of 6\,arcsec$^2$, with 60\% of the emission due to the OF1 outflow.
The results are in Figs.\,\ref{H2O-final-2in1} and \ref{blue}, Tables\,\ref{tableappendix1} and\,\ref{tableappendix2}. 
To provide an estimate on modelling uncertainties, we  added the results of the radiative transfer computed in stationary state instead of statistical equilibrium \citep[see][for details]{Gusdorf11}, and for a slightly  slower shock model to account for the positional discrepancy between SiO and H$_2$O observations. 
The o-H$_2$O line at 1153\,GHz  is dramatically over-predicted by all models. However, this transition is masing in our LVG calculations \citep[and in RADEX,][]{2007A&A...468..627V} and therefore predictions  are not reliable. 
Three high-lying transitions are nicely reproduced in terms of maximum brightness temperature and
integrated intensity in both the red- and blue-shifted component, although with a smaller area for the blue shifted case, 3\,arcsec$^2$. Estimates of the SiO lines with this area are still compatible
with the observations, and there is no other constraint on the area of the blue lobe since H$_2$ is not detected. 
The low-energy lines (p-H$_2$O
at 1113 and 988\,GHz, Table\,\ref{table1}) are over-predicted by the model. Three
explanations might be invoked to explain this discrepancy. First, these lines could be partly self-absorbed even at the
high-velocities used in our analysis. This could be true for the
1113\,GHz red-wing, as suggested by the sharp absorption at
$+6$\,km\,s$^{-1}$, directly at the edge of the velocity range used for
the shock analysis, and by the comparison with H$_2^{18}$O $1_{11}-0_{00}$
detected in emission at
lower velocities than H$_2$O. However, there is no evidence for
self-absorption in the 988\,GHz line.
The optical thickness of these lines might also explain the discrepancy between models and data (non-local radiative transfer might affect their emissivity more than in the other lines). But the most convincing argument is that H$_2$O  could  be dissociated in the quiescent parts of the shock, affecting the transitions that are most likely to emit in these regions. In this case, one should detect emission from the most abundant photo-dissociation products, namely OH and O \citep{1988ASSL..146...49V}. Future observations with SOFIA might help to support this scenario. Refined shock-codes including  effects of  radiation fields are also needed to address this question.

If we accept that the SiO model also fits the H$_2$O emission,  we
 can infer the column
density and the mass of H$_2$O in OF1  because the column density is self-consistently computed
in our shock model, and  we have constraints on the area of the emission
region. Whether we adopt an age of 500 or 1000\,yr (Paper\,II), the maximum H$_2$O fractional
abundance with respect to $n_H$ in the shocked layer is $x_{\rm
  H_2O}\simeq1.4\,$10$^{-4}$. The corresponding water abundance 
in fractional column density units is $2.5\,10^{-5}$ 
for a dynamical age of 500\,yr, and $1.2\,10^{-5}$ for an age of 1000\,yr (see Appendix\,\ref{sec:twapomm}).
The corresponding column density
over the shock layer is $N_{\rm H_2O} =
1.2\,$10$^{18}$\,cm$^{-2}$, almost a factor of two  higher than  the lower limit ($5.3\,10^{17}$\,cm$^{-2}$) found in Sect.\,\ref{obs_res} based on crude assumptions.
For an area of 6\,arcsec$^2$ for the red-lobe and of 3\,arcsec$^2$ for the blue one, at 1\,kpc distance this column
density corresponds to a shocked water mass of 
3.8$\,$10$^{-5}\,M_\odot$, or 
12.5\,$M_\oplus$.  

In our model, the maximum of the local H$_2$O density is attained 45\,yr after the temperature peak. The highest value is a result of  sputtering of the ices in the grain mantles, and of high-temperature chemistry. Because the sputtering is simultaneous to the temperature rise, 45\,yr is the time scale for the high-temperature chemistry under these shock conditions.
Given the small O$_2$ abundance measured in dense cold molecular clouds, water is mainly formed via the sputtering of grain mantles, for which standard models predict a total release of material towards the gas phase above a shock velocity threshold of 20--25 km\,s$^{-1}$ \citep[e.g.,][]{1983ApJ...264..485D,1994MNRAS.268..724F}. 
 Since both shock velocities used in our analysis are well above the  threshold shock speed for water, the derived H$_2$O abundance does not change significantly at  $\varv_{\rm s}$=30\,km\,s$^{-1}$. 

\section{Discussion and conclusions}\label{dis}

The SiO(8--7) and H$_2$O  profiles (in particular that of the 1113\,GHz line) suggest a common origin of the H$_2$O and SiO emission in IRAS\,17233. This result is based on emission at high velocities and  
is different from the findings that SiO and H$_2$O do not trace the same gas in  molecular outflows from low-mass YSOs at low-velocities and/or in low-energy lines 
\citep{2012A&A...538A..45S,2013A&A...549A..16N}. However, an excellent match between SiO and H$_2$O profiles  is found in  other sources at high velocities \citep{2012ApJ...757L..25L}.

With the limitations previously discussed, we find that the shock
parameters of OF1 are comparable with those found for low-mass
protostars with a higher pre-shock density. The derived
water abundance is compatible with values of other molecular
outflows  \citep[e.g.,][]{2010A&A...521L..28E,2012A&A...540A..84H}. While often
measurements of H$_2$O abundances have large uncertainties because the
H$_2$ column density is inferred from observations of CO 
or from models \citep[for a compilation of sources, abundances and methods, see][]{2013ChRv..113.9043V}, the value inferred in our analysis is consistently derived, as the H$_2$O and  H$_2$  column densities are outcomes of the same  model. Moreover, the estimated H$_2$O column density matches the data.
Although photo-dissociation probably affects the low-energy H$_2$O lines, simple C-shocks models can be used to model higher-energy transitions. 
The inclusion of photo-dissociation in our models is work in
progress in a larger framework of studying the effect of an intense UV
field on shocks.

Estimates of H$_2$O mass are not easily found in the literature. 
\citet{2014A&A...561A.120B} modelled water emission in L1157-B1 through J- and C-type
shocks. Their H$_2$O column densities derived over the whole line
profiles translate in to masses in the range 0.009--0.125\,$M_\oplus$
for a hot component of 2\arcsec--5\arcsec size and
$<(0.7-1.5)\,10^{-3}\,M_\oplus$ for a warm component with a size of
$\le$10\arcsec. Our estimate of  12.5\,$M_\oplus$ for the H$_2$O mass of OF1  therefore seems to be compatible with previous results.

In summary, we presented the first estimate of the abundance of water in an outflow driven by a massive YSOs based on a self-consistent shock model  of water and SiO transitions. We inferred a water abundance in fractional column density units between $1.2\,10^{-5}$ and $2.5\,10^{-5}$, which is an average value of the water abundance over the shock layer. 
Additionally, our model indicates that the maximum fractional abundance of water locally reached in the layer is $10^{-4}$. Finally, we inferred the water mass of the OF1 outflow to be 12.5\,$M_\oplus$.

\begin{acknowledgements}
{\it Herschel} is an ESA space observatory with science instruments provided
by European-led Principal Investigator consortia and with important
participation from NASA. 
HIFI has been designed and built by a consortium of
 institutes and university departments from across Europe, Canada and the
 United States under the leadership of SRON Netherlands Institute for Space
 Research, Groningen, The Netherlands and with major contributions from
 Germany, France and the US. Consortium members are: Canada: CSA,
 U.Waterloo; France: CESR, LAB, LERMA, IRAM; Germany: KOSMA,
 MPIfR, MPS; Ireland, NUI Maynooth; Italy: ASI, IFSI-INAF, Osservatorio
 Astrofisico di Arcetri- INAF; Netherlands: SRON, TUD; Poland: CAMK, CBK;
 Spain: Observatorio Astron{\'o}mico Nacional (IGN), Centro de
 Astrobiolog{\'i}a
 (CSIC-INTA). Sweden: Chalmers University of Technology - MC2, RSS $\&$
 GARD; Onsala Space Observatory; Swedish National Space Board, Stockholm
 University - Stockholm Observatory; Switzerland: ETH Zurich, FHNW; USA:
 Caltech, JPL, NHSC.

A.\,G. acknowledges support by the grant ANR-09-BLAN-0231-01 from the \textit{French Agence Nationale de la Recherche} as part of the SCHISM project. T.\,Cs. is funded by the ERC Advanced Investigator Grant GLOSTAR (247078). A.\,G. acknowledges useful discussions with C.\,Vastel and A.\,Coutens.
\end{acknowledgements}

\Online

\begin{appendix}

\section{Water abundance problem: the point of view of observers and modellers}\label{sec:twapomm}

The goal of this appendix is to clarify  the possible confusion of  the meaning of "water abundance" between the observing and modelling communities. The rigorous comparison of observations to models requires the knowledge of constraints such as the length/age of the shock, as this section discusses now.
We base this discussion on the
model used to fit both the SiO and H$_2$O emission in the OF1
shock region of IRAS 17233--3606 with the
following input parameters:\,pre-shock density $n_{\rm H}
=$\,10$^6$\,cm$^{-3}$, shock velocity $\varv_{\rm s}\,=$\,32\,km\,s$^{-1}$,
and magnetic field strength (perpendicular to the shock direction) $B
=$\,1\,mG. Whether the radiative transfer of water is calculated along the shock equations in the model (so-called \lq s-s' in Fig. 3, \lq DRF' in Tables\,\ref{tableappendix1}--\ref{tableappendix4}) or a posteriori from the outputs of the shock model (\lq s-e' in Fig. 3, \lq AGU' in Tables\,\ref{tableappendix1}--\ref{tableappendix4}) does not change the thermal profile of the shock layer, nor the associated water abundances (e.g. Gusdorf et al. 2011). Everything stated in this appendix is therefore applicable to both \lq s-s' and \lq s-e' models.

In one-dimensional, stationary shock models (e.g., this work, \citealt{Gusdorf11,1983ApJ...264..485D,Neufeld96,Flower10}) the
physical and chemical conditions are self-consistently calculated at
each point of a shocked layer. The end product is a collection of
physical (temperature, velocity, density) and chemical (abundances)
quantities obtained at each point of the shocked layer. The position
of each point is marked by a distance parameter with respect
to a origin typically located in the pre-shock region. The
position of the last point in the post-shock region then corresponds
to the shock width. Typically these shock models are used in a face-on
configuration, so that the width one refers to is along the
line-of-sight direction. 
Alternatively, the position of a point in
the shock layer can be expressed through a time parameter: the time
parameter for the last point in the post-shock region then corresponds
to the flight time that a particle needs to flow through the total
width of the shock. The correspondence between the time and distance
parameters related to a neutral particle ($t_{\rm n}$ and $z$) is
hence given by $t_{\rm n} = \int (1/\varv_{\rm n})\,dz$, where
$\varv_{\rm n}$ is the particle velocity. While the shock width cannot be constrained by observations,
an upper limit to the flow time is given by the dynamical age,
which is  inferred from mapped observations of spectrally resolved
lines.

\begin{figure}
\centering
\includegraphics[width=0.45\textwidth]{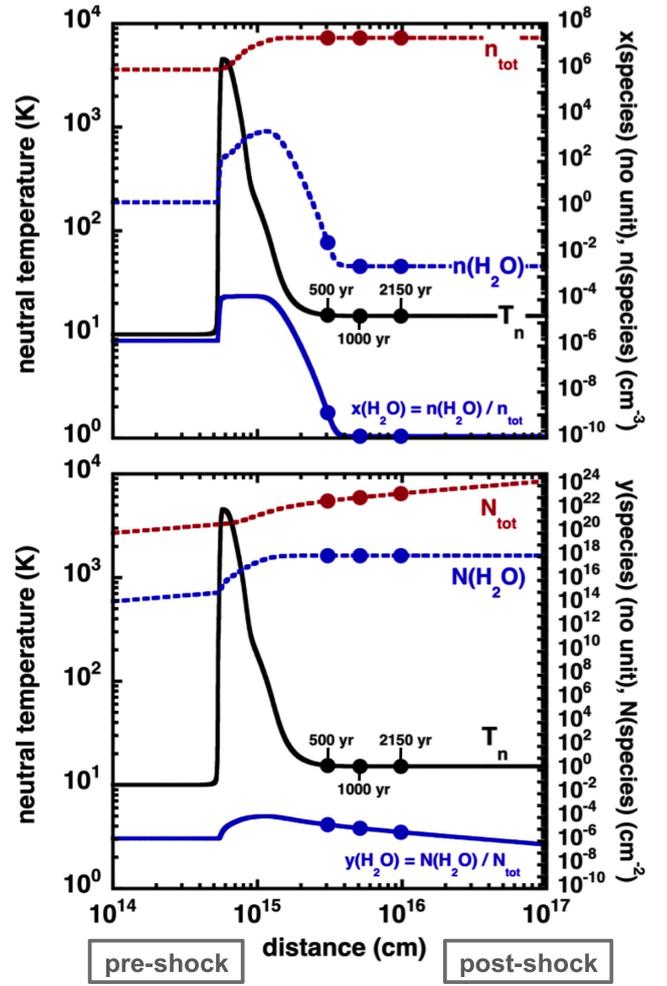}
\caption{\textit{Upper panel:}\,the neutral temperature (black curve), total density (red dashed curve), water density (blue dashed curve), and fractional density (blue continuous curve). The so-called fractional density is the water density over the total density, locally defined at each point of the shock. \textit{Lower panel:}\,the neutral temperature (black curve), total column density (red dashed curve), water column density (blue dashed curve), and fractional column density (blue continuous curve). The so-called fractional column density is the water column density over the total column density. The column density (in cm$^{-2}$) is the integral of the local density (in cm$^{-3}$) along the shock width (in cm). In both panels, the three points labelled on each curve correspond to the distance parameter of $3.1\,10^{15}, 5.15\,10^{15}, 10^{16}$\,cm, or to time parameters values of 500, 1000, and 2150\,yr.}
\label{appendix-fig}
\end{figure}
   
Figure\,\ref{appendix-fig}  shows for this
model the variation of the temperature of the neutral particles (K),
as well as those of the water and total local densities ($n$(H$_2$O) and $n_{\rm{tot}}$ in cm$^{-3}$) and their ratio
$x$(H$_{2}$O)\,=\,$n$(H$_2$O)/$n_{\rm tot}$ in the shock layer versus
the distance parameter. To illustrate the relation between time and
distance parameters through the shock layer, we have marked three
points on each curve:\,$3.1\,10^{15},
5.15\times10^{15}, 10^{16}$\,cm, which correspond to 500, 1000, and
2150\,yr, in our model. In our case, the highest value for the time parameter is
constrained by  the dynamical shock age of OF1, 
500--1000\,yr. Water abundance is often defined by modellers  as the
\textit{maximum fractional local abundance} of water through the shock layer,
that is, between the pre-shock region before the temperature rise and the
maximum shock age ($x$(H$_2$O)$_{\rm max} = 1.4\,10^{-4}$ for our model, 
top panel of Fig.\,\ref{appendix-fig}).
On the other hand, local quantities cannot be accessed through
observations. Integrated quantities (against the width of the shock
layer along the line of sight) such as column densities are measured
by observers. Generally, \lq observational water abundances' are hence
given in fractional column density units, that is, the ratio of the
water column density divided by the total column density. This ratio
is different the maximum fractional abundance of
water that is generally provided and used by modellers. The difference
between the two values is illustrated by comparing the upper panel of
Figure\,\ref{appendix-fig} with its lower panel, which shows the evolution
of the water and total column densities, $N_{\rm H_2O}$ and $N_{\rm
  tot}$, and of their ratio $y$(H$_{2}$O)\,=\,$N$(H$_2$O)/$N_{\rm
  tot}$. In the modellers' view, referring to the distance parameter as
\lq z', these column densities are defined by
\begin{equation}
N{\rm (H_2O)} \rm{[cm^{-2}]} = \int_{0}^{z_{\rm max}} \it{n}{\rm (H_2O)} \rm{[cm^{-3}]}\, \it{dz},
\end{equation}
\begin{equation}
N_{\rm tot} \rm{[cm^{-2}]} = \int_{0}^{z_{\rm max}} \it{n}_{\rm tot} \rm{[cm^{-3}]}\, \it{dz},
\end{equation}

where $z_{\rm max}$ is the total shock width, that is, the distance corresponding to the maximum value of the time parameter. In our case, the value of the fractional column density of water can be read in the bottom panel of Fig.\,\ref{appendix-fig}:\,$y$(H$_2$O)$ = 2.5\times10^{-5}$ (if the adopted dynamical age is 500 yr), $ = 1.2\,10^{-5}$ (if the adopted dynamical age is 1000 yr). We note that this value is about an order of magnitude lower than the maximum fractional abundance of water reached in the same shock layer.

We note that the decrease in the $y$(H$_2$O) curve is artificial and only due to the 1D nature of the model. Indeed, in the post-shock region, the total density of the gas is conserved (because it cannot escape sideways, for instance like in the case of a bow-shock), while the gas-phase water density decreases until all water molecules re-condensate on the interstellar grains because of the temperature decrease. The total column density hence increases (lower panel of Fig.\,\ref{appendix-fig}), while the water column density is constant, resulting in a decrease of the water column density ratio with the distance or time parameter. It is therefore essential to have a measurement of the dynamical time scale to stop the calculation at a realistic time to obtain a fractional column density of water  as realistic as possible.

\section{Additional tables and figures}
\begin{table*}
\caption{Summary of the observations.}
\label{table1}      
\centering          
\begin{tabular}{l r r c c c c c c c}     
\hline \hline
Line&\multicolumn{1}{c}{$\nu^1$}&\multicolumn{1}{c}{$E_{\rm{up}}^1$}&\multicolumn{1}{c}{Beam$^2$}&\multicolumn{1}{c}{$\eta_{\rm{mb}}^2$}&\multicolumn{1}{c}{$T_{\rm
    {sys}}$}&\multicolumn{1}{c}{$\delta
  \rm{v}$}&\multicolumn{1}{c}{r.m.s.}&\multicolumn{1}{c}{OBSIDs}&\multicolumn{1}{c}{mode$^3$}\\ 
&(GHz)&(K)&(\arcsec)&&(K)&(km\,s$^{-1}$)&(K)\\ \hline
$p$-H$_2$O $4_{22}-3_{13}$&1207.639&454.5&17.6&0.64&1063&0.12&0.21&1342242862&DBS\\ 
$o$-H$_2$O $3_{21}-3_{12}$&1162.912&305.4&18.2&0.64&850&0.13&0.18&1342242863&DBS\\ 
$o$-H$_2$O $3_{12}-2_{21}$&1153.127&249.5&18.3&0.64&836&0.13&0.18&1342242863&DBS\\ 
$p$-H$_2$O $1_{11}-0_{00}$&1113.343&53.5&19.0&0.74&389&0.10&0.13&1342266536&DBS\\ 
$p$-H$_2^{18}$O $1_{11}-0_{00}$&1101.698&52.9&19.0&0.74&389&0.10&0.13&1342266536&DBS\\
 $p$-H$_2$O $2_{02}-1_{11}$&987.927&100.9&21.5&0.74&333&0.15&0.15&1342242875&DBS\\
 $p$-H$_2$O $2_{11}-2_{02}$&752.033&137.0&28.2&0.74&187&0.20&0.20&1342266457&DBS\\ 
\hline
\end{tabular}
\tablefoot{\tablefoottext{1}{\citet{pickett_JMolSpectrosc_60_883_1998}.}\tablefoottext{2}{Half-power beam width and main beam efficiency from \citet{2012A&A...537A..17R}.}\tablefoottext{3}{DBS stands for dual beam switch mode}.}
\end{table*}

\begin{table*}
\caption{Observed and modelled maximum line temperatures ($T^{\rm max}$, K) for the red lobe.}
\label{tableappendix1}      
\centering       
\begin{tabular}{r c c c c c c c c} 
\hline
\hline
$\nu$ & $E_{\rm up}$ & Beam & $FF^{-1}$\tablefootmark{(1)} & $T_{\rm obs}^{\rm max}$ & $T_{\rm obs, corr}^{\rm max}$\tablefootmark{(2)}& $T_{\rm AGU 32}^{\rm max}$\tablefootmark{(3)}& $T_{\rm AGU 30}^{\rm max}$\tablefootmark{(4)} & $T_{\rm DRF 32}^{\rm max}$\tablefootmark{(5)}  \\	
(GHz) & (K) & (\arcsec) & (no unit) & (K) & (K) & (K) & (K) & (K)  \\
\hline
1113 & 53.4 & 19.1 & 48.5 & 2.2 & 64.0 & 169.8 & 159.1 & 141.5  \\
988 & 100.8 & 21.5 & 61.4 & 3.3 & 121.6 & 229.3 & 215.9 & 192.6  \\
752 & 136.9 & 28.2 & 105.1 & 3.3 & 208.1 & 204.3 & 195.1 & 139.8  \\ 
1153 & 249.3 & 18.4 & 45.3 & 2.9 & 78.8 & 735.9 & 676.1 & 356.9  \\
1163 & 305.3 & 18.2 & 44.5 & 3.5 & 93.5 & 79.4 & 61.0 & 58.1  \\
1208 & 454.3 & 17.6 & 41.4 & 1.0 & 24.8 & 35.4 & 20.9 & 21.7  \\
\hline
\hline
\end{tabular}
\tablefoot{\tablefoottext{1}{inverse of the beam filling factor at each frequency considering an emitting area of 6\,arcsec$^2$.}
\tablefoottext{2}{Observed maximum temperature corrected for filling factor and 60\% contribution of OF1.}
\tablefoottext{3}{Modelled maximum temperature following \citet{Gusdorf11} with $\varv_{\rm s}=32$\,km\,s$^{-1}$.}
\tablefoottext{4}{Modelled maximum temperature following \citet{Gusdorf11} with $\varv_{\rm s}=30$\,km\,s$^{-1}$.}
\tablefoottext{5}{Modelled maximum temperature following \citet{Flower10} with $\varv_{\rm s}=32$\,km\,s$^{-1}$}.}
\end{table*}

\begin{table*}
\caption{Observed and modelled integrated intensities ($\int T d\varv$, K km s$^{-1}$) for the red lobe.}
\label{tableappendix2}      
\centering       
\begin{tabular}{r c c c c c c c c} 
\hline
\hline
$\nu$ & $E_{\rm up}$ & Beam & $FF^{-1}$\tablefootmark{(1)} & [$\int T d\varv$]$_{\rm obs}$ & [$\int T d\varv$]$_{\rm corr}$\tablefootmark{(2)} & [$\int T d\varv$]$_{\rm AGU 32}$ & [$\int T d\varv$]$_{\rm AGU 30}$ & [$\int T d\varv$]$_{\rm DRF 32}$ \\	
(GHz) & (K) & (\arcsec) & (no unit) & K km s$^{-1}$ & K km s$^{-1}$ & K km s$^{-1}$ & K km s$^{-1}$ & K km s$^{-1}$ \\
\hline
1113 & 53.4 & 19.1 & 48.5 & 32.0 & 931.42 & 1743.0 & 1547.0 & 2006.7 \\
988 & 100.8 & 21.5 & 61.4 & 42.4 & 1562.1 & 2914.0 & 2606.0 & 2810.5 \\
752 & 136.9 & 28.2 & 105.1 & 36.6 & 2307.5 & 2112.0 & 1894.0 & 1976.3 \\ 
1153 & 249.3 & 18.4 & 45.3 & 38.4 & 1043.7 & 5188.0 & 4512.0 & 3366.6 \\
1163 & 305.3 & 18.2 & 44.5 & 21.1 & 563.4 & 534.6 & 462.5 & 752.1 \\
1208 & 454.3 & 17.6 & 41.4 & 13.0 & 322.9 & 194.2 & 158.0 & 223.8 \\
\hline
\hline

\end{tabular}
\tablefoot{\tablefoottext{1}{inverse of the beam filling factor at each frequency considering an emitting area of 6\,arcsec$^2$.}
\tablefoottext{2}{Observed integrated intensity corrected for filling factor and 60\% contribution of OF1.}
\tablefoottext{3}{Modelled integrated intensity following \citet{Gusdorf11} with $\varv_{\rm s}=32$\,km\,s$^{-1}$.}
\tablefoottext{4}{Modelled integrated intensity following \citet{Gusdorf11} with $\varv_{\rm s}=30$\,km\,s$^{-1}$.}
\tablefoottext{5}{Modelled integrated intensity following \citet{Flower10} with $\varv_{\rm s}=32$\,km\,s$^{-1}$}.}
\end{table*}

\begin{table*}
\caption{Observed and modelled maximum line temperatures ($T^{\rm max}$, K) for the blue lobe.}
\label{tableappendix3}      
\centering       
\begin{tabular}{r c c c c c c c c} 
\hline
\hline
$\nu$ & $E_{\rm up}$ & Beam & $FF^{-1}$\tablefootmark{(1)} & $T_{\rm obs}^{\rm max}$ & $T_{\rm obs, corr}^{\rm max}$\tablefootmark{(2)} & $T_{\rm AGU 32}^{\rm max}$\tablefootmark{(3)} & $T_{\rm AGU 30}^{\rm max}$\tablefootmark{(4)} & $T_{\rm DRF 32}^{\rm max}$\tablefootmark{(5)}  \\	
(GHz) & (K) & (\arcsec) & (no unit) & (K) & (K) & (K) & (K) & (K)  \\
\hline
988 & 100.8 &  21.5 & 121.8 & 1.1 & 83.3 & 229.3 & 215.9 & 192.6  \\
1153 & 249.3 & 18.4 & 89.5 & 1.2 & 66.3 & 735.9 & 676.1 & 356.9  \\
1163 & 305.3 & 18.2 & 88.0 & 1.0 & 53.9 & 79.4 & 61.0 & 58.1  \\
1208 & 454.3 & 17.6 & 81.8 & 0.5 & 23.6 & 35.4 & 20.9 & 21.7  \\
\hline
\hline
\end{tabular}
\tablefoot{\tablefoottext{1}{inverse of the beam filling factor at each frequency considering an emitting area of 3\,arcsec$^2$.}
\tablefoottext{2}{Observed maximum temperature corrected for filling factor and 60\% contribution of OF1.}
\tablefoottext{3}{Modelled maximum temperature following \citet{Gusdorf11} with $\varv_{\rm s}=32$\,km\,s$^{-1}$.}
\tablefoottext{4}{Modelled maximum temperature following  \citet{Gusdorf11} with $\varv_{\rm s}=30$\,km\,s$^{-1}$.}
\tablefoottext{5}{Modelled maximum temperature following  \citet{Flower10} with $\varv_{\rm s}=32$\,km\,s$^{-1}$}.}
\end{table*}

\begin{table*}
\caption{Observed and modelled integrated intensities ($\int T d\varv$, K km s$^{-1}$) for the blue lobe.}
\label{tableappendix4}      
\centering       
\begin{tabular}{r c c c c c c c c} 
\hline
\hline
$\nu$ & $E_{\rm up}$ & Beam & $FF^{-1}$\tablefootmark{(1)} & [$\int T d\varv$]$_{\rm obs}$ & [$\int T d\varv$]$_{\rm corr}$\tablefootmark{(2)} & [$\int T d\varv$]$_{\rm AGU 32}$\tablefootmark{(3)} & [$\int T d\varv$]$_{\rm AGU 30}$\tablefootmark{(4)} & [$\int T d\varv$]$_{\rm DRF 32}$\tablefootmark{(5)} \\	
(GHz) & (K) & (\arcsec) & (no unit) & K km s$^{-1}$ & K km s$^{-1}$ & K km s$^{-1}$ & K km s$^{-1}$ & K km s$^{-1}$ \\
\hline
988 & 100.8 & 21.5 & 121.8  & 9.8 & 714.9 & 2914.0 & 2606.0 & 2810.5 \\
1153 & 249.3 & 18.4 &89.5   & 12.2 &653.3 & 5188.0 & 4512.0 & 3366.6 \\
1163 & 305.3 & 18.2 &88.0   & 7.9 & 415.1 & 534.6 & 462.5 & 752.1 \\
1208 & 454.3 & 17.6 &81.8   & 3.7 & 183.1 & 194.2 & 158.0 & 223.8 \\
\hline
\hline

\end{tabular}
\tablefoot{\tablefoottext{1}{inverse of the beam filling factor at each frequency considering an emitting area of 3\,arcsec$^2$.}
\tablefoottext{2}{Observed integrated intensity corrected for filling factor and 60\% contribution of OF1.}
\tablefoottext{3}{Modelled integrated intensity following \citet{Gusdorf11} with $\varv_{\rm s}=32$\,km\,s$^{-1}$.}
\tablefoottext{4}{Modelled integrated intensity following with $\varv_{\rm s}=30$\,km\,s$^{-1}$}
\tablefoottext{5}{Modelled integrated intensity following \citet{Flower10} with $\varv_{\rm s}=32$\,km\,s$^{-1}$}.}
\end{table*}

\begin{figure}
\centering
\includegraphics[width=0.45\textwidth]{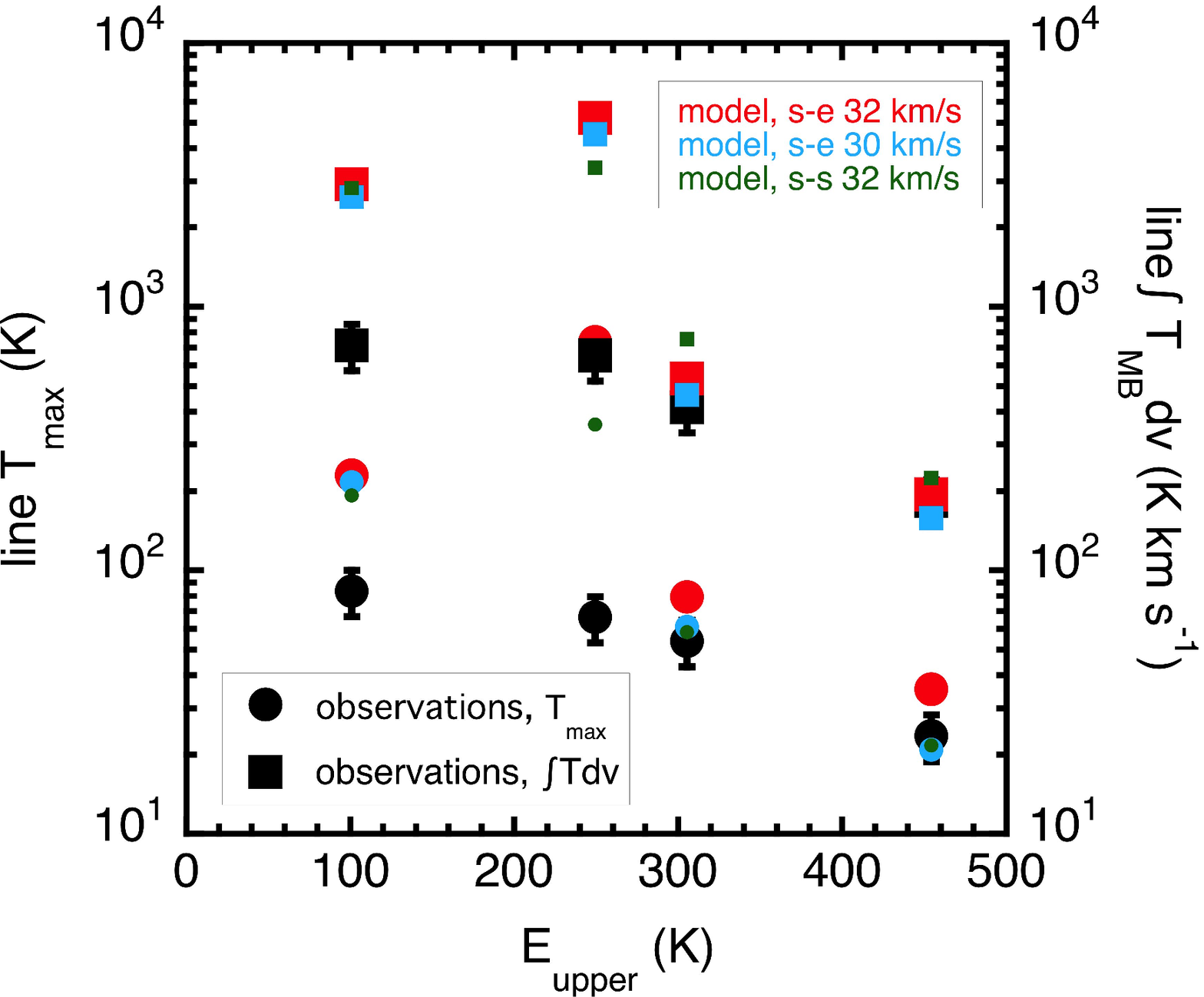}
\caption{Observed and modelled maximum brightness temperatures (circles), and integrated intensities (squares) for the blue-shifted emission. Data points (in black) are corrected for an emission region of 3\,arcsec$^2$ and for 60\% of the emission due to OF1. Errorbars are $\pm$20\% of the observed value. Three models are shown: the model of Paper\,II with level populations in statistical equilibrium (\lq s-e' in red) with $\varv_{\rm s}=32$\,km\,s$^{-1}$, one with a slower shock velocity ($\varv_{\rm s}=30$\,km\,s$^{-1}$, blue), and a model in stationary-state (\lq s-s' in green).}
\label{blue}
\end{figure}

\end{appendix}

\end{document}